\documentclass[]{IAC_style}

\begin{document}

\IACpaperyear{2024} 
\IACpapernumber{IAC-24-D1-3-1-x84065} 
\IAClocation{Milan, Italy} 
\IACdate{14-18 October 2024} 

\IACcopyrightB{'copyright holder'}

\title{Advancing lunar exploration through virtual reality simulations:  
a framework for future human missions}

\IACauthor{Giacomo Franchini$^{\orcidlink{0009-0009-5641-8346}}$}{1}{1}
\IACauthor{Brenno Tuberga}{1}{0}
\IACauthor{Marcello Chiaberge$^{\orcidlink{0000-0002-1921-0126}}$}{1}{0}

\IACauthoraffiliation{Department of Electronics and Telecommunications, Polytechnic of Turin, Corso Duca degli Abruzzi, 24, 10129 Torino, Italy \normalfont{E-mail:~\authormail{giacomo.franchini@polito.it}, ~\authormail{brenno.turberga@polito.it}, ~\authormail{marcello.chiaberge@polito.it}}}

\abstract{
In an era marked by renewed interest in lunar exploration and the prospect of establishing a sustainable human presence on the Moon, innovative approaches supporting mission preparation and astronaut training are imperative. To this end, the advancements in Virtual Reality (VR) technology offer a promising avenue to simulate and optimize future human missions to the Moon. Through VR simulations, tests can be performed quickly, with different environment parameters and a human-centered perspective can be maintained throughout the experiments. This paper presents a comprehensive framework that harnesses VR simulations to replicate the challenges and opportunities of lunar exploration, aiming to enhance astronaut readiness and mission success. Multiple environments with physical and visual characteristics that reflect those found in interesting Moon regions have been modeled and integrated into simulations based on the Unity graphical engine. We exploit VR to allow the user to fully immerse in the simulations and interact with assets in the same way as in real contexts. Different scenarios have been replicated, from upcoming exploration missions where it is possible to deploy scientific payloads, collect samples, and traverse the surrounding environment, to long-term habitation in a futuristic lunar base, performing everyday activities. Moreover, our framework allows us to simulate human-robot collaboration and surveillance directly displaying sensor readings and scheduled tasks of autonomous agents which will be part of future hybrid missions, leveraging the ROS2-Unity bridge. Thus, the entire project can be summarized as a desire to define cornerstones for human-machine design and interaction, astronaut training, and learning of potential weak points in the context of future lunar missions, through targeted operations in a variety of contexts as close to reality as possible.
}

\maketitle

\section*{Acronyms}
Augmented Reality (AR)

China National Space Administration (CNSA)

Deep Learning (DL)

Extra Vehicular Activity (EVA)

Human Centered Design (HCD)

In Situ Resource Utilization (ISRU)

International Space Station (ISS)

Positioning, Navigation and Timing (PNT)

Robot Operating System 2 (ROS2)

Transmission Control Protocol (TCP)

Unified Robot Description Format (URDF)

Virtual Reality (VR)

\section{Introduction}
More than five decades after the last human mission to the Moon, we are experiencing an era of renewed interest in lunar exploration from government agencies and private players. Many factors are driving this second race to our satellite: from the possibility of accessing potentially useful lunar resources to the motivation of building a permanent base on its surface, setting the Moon as a gateway for future deep space missions. As a first step in this direction, National Aeronautics and Space Administration is developing the Artemis program, planning to land again astronauts on the Moon by 2026. The agency is also collaborating with private companies through the Commercial Lunar Payload Services, to deliver scientific payload and technological demonstrators. On the other end, with the multiple Chang’e missions, CNSA advanced its lunar exploration program, allowing the deployment of robotic missions to the Moon surface and the collect samples from its far side and bring them back to Earth, with the agency's long term goal of establish a permanent lunar base near the South Pole. 

Although steps are being taken in this direction, a lot of effort is still needed in the prospect of establishing a sustainable human presence on the Moon. Today's missions principally focus on scientific exploration, but before being able to permanently stay on the lunar surface, a lot of infrastructure is needed, such as a satellite constellation for precise PNT and systems for managing ISRU, construction and logistics. 

\subsection{Related work}
In view of a long-term human mission, particular importance must be given to astronaut's preparation and training. In this context, the latest developments in simulations and VR technologies could be exploited. Through simulations, different mission scenarios and environment parameters can be replicated and tested, while using VR a human-centered point of view is maintained during experiments. This is confirmed by \cite{10503238}, where the authors survey VR applications for astronaut training. They illustrate how a training approach based on realistic simulations is highly versatile, customizable, and efficient in terms of safety, costs, and time savings. Again, regarding task preparation to be performed by astronauts during a mission, \cite{Garcia} presents a framework where VR is integrated with hardware-in-the-loop simulations that allow the trainee to experience physical feedback while executing the task. Various operation scenarios are simulated, such as EVA activities, mass handling, and collaboration with robotics procedures, but all are relative to missions to be performed on the ISS.        
The concept of leveraging VR to enhance the engineering of early-stage lunar mission prototypes is well described in \cite{Testi, Nilsson}. In particular, \cite{Testi} focuses on the development of an iterative design process of lunar habitats in which VR is integrated to obtain valuable feedback on design choices directly from a human perspective. In \cite{Nilsson}, instead, the authors modeled a lunar exploration scenario and showed how VR facilitates the user-centered design of tools and assets to be manipulated by astronauts during a future mission. Moreover, \cite{McHenry} implements a framework aimed at helping astronauts navigate the lunar environment during their missions. It leverages Deep Learning (DL) and Augmented Reality (AR) to show in real-time waypoints to be followed, potential key points, and dangerous zones. The framework can be trained based on data collected from robotic missions at the site, in preparation for human landing.
An example of a simulation of human-robot collaboration is shown in \cite{curlin2022virtualrealitydigitaltwin}. Here a rover Digital Twin is spawned in a simulated lunar environment and can be teleoperated by astronauts during training sessions to replicate and escape from possible rover failure situations.

\subsection{Paper outline}     
In this paper, we introduce a VR-based framework that simulates lunar exploration challenges to enhance astronaut readiness and mission success. Using Unity game engine, we model environments resembling key Moon regions, enabling the replication of interesting scenarios with an immersive user experience. The framework also supports human-robot collaboration, integrating sensor data and autonomous task management. This paper is structured as follows: in Section 2 we present in detail the methodology followed for the implementation of our framework. In Section 3, three different lunar mission scenarios are presented highlighting the key role of VR and HCD, while in Section 4 we describe the results obtained and discuss future extensions to our framework.  

\section{Material and methods}
In this section, we illustrate the methodologies and technologies used for the implementation of our astronaut's training framework. First, we describe how assets are modeled and how we simulate lunar environments, together with VR integration in the simulations. Then, we focus on the connection between our framework and the ROS2 environment to perform robotic tasks and human-robot collaboration.    

\subsection{Simulations}
Simulations are based on the popular Unity Game Engine \cite{unity}. Unity allows developers to create 2D, 3D, and VR experiences across various platforms, including consoles, mobile devices, and PCs. It offers a robust suite of tools for designing interactive environments, physics simulations, and complex animations. We choose Unity over other game engines mainly for its ease of use, thanks to its user interface and vast C\# APIs, for its versatility and compatibility with VR hardware available in our lab.
Simulations in Unity are organized into projects, which contain all assets, scripts, and dependencies needed. Inside the project, the user can define the scene to be simulated, import the desired assets, and organize them in a hierarchical way. Unity also features prefabs: reusable and reconfigurable assets acting as a template. These are particularly useful for example to spawn two equivalent exploration rovers with different onboard sensor configurations. Moreover, Unity natively imports Blender \cite{blender} objects thanks to the Blender FBX exporter. This feature has been a key factor during the simulation engine choice since basically all assets and materials in our framework have been modeled in Blender.            

\subsection{Virtual reality}
For realizing VR experiences we employ the Meta Quest 3 headset. Featuring two LCD displays with a single resolution of 2064×2208p and a horizontal field of view of 110°, the headset allows users to fully immerse in the simulated environments. The headset ships with two dedicated Meta Touch Plus controllers, employed for manipulation tasks. The integration of Meta Quest 3 with Unity is performed thanks to the Meta XR All-in-One SDK \cite{meta_xr_sdk} packages suite. This is a wrapper for single packages consenting developers to implement desired VR applications, exploiting the headset capabilities such as body pose detection, hands and controllers interaction for object grab and interface with the environment, haptic and audio feedback. Moreover, we exploited the Meta Movements SDK \cite{meta_movements_sdk} for realistically animate astronaut avatars, based on feedback from headset and controllers. This kit allows us to connect user movements in the real world to the simulated rigged avatar in Unity. 

\subsection{Human-robot collaboration} \label{sub:Human-robot collaboration}
As it is one of the most popular development environments for robotic applications, we employed ROS2 \cite{ROS2} to integrate autonomous and collaborative tasks of robotic agents in our simulations. ROS2 is open source and provides a set of useful libraries to implement a multitude of robot behaviors. Moreover, the framework is fully integrated with Unity thanks to the set of tools provided by Unity Robotics Hub \cite{unity_ros2}: these include packages for setting up the communication layer to exchange messages over topics and performing service calls, as typical in ROS2, for importing robot models from URDF format and for simulating various sensors. On the Unity side, the main components to set the communication are the two Unity plugins MessageGeneration and ROSConnection. The first implements the generation of dedicated C\# classes for every supported ROS2 message type together with its relative serialization, allowing their usage in Unity scripts, while the second is the input/output point for serialized messages from and to the ROS2 network. On the ROS2 side, a TCP endpoint component manages message traffic between specific application nodes and the engine. 

\begin{figure}[t!]
\centering
\includegraphics[width=\columnwidth]{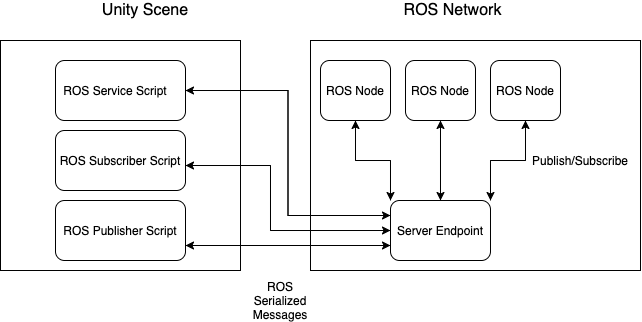}
\caption{Unity - ROS2 communication diagram. From \cite{unity_ros2}.}
\label{fig:unity_ros_comm}
\end{figure}

The communication diagram is illustrated in Figure \ref{fig:unity_ros_comm}.

\section{Astronauts training simulation framework} 
We will now explore in detail our astronaut training framework. In particular, three hypothetical mission scenarios will be discussed: in the first, the user can immerse their-self in a future lunar base concept. In the second, a scene is set up to train sample collection tasks. Lastly, an example of collaboration between humans and robots during the exploration of unknown Moon regions is shown.         
\subsection{Lunar base concept}
\begin{figure}[h!]
\centering
\includegraphics[width=\columnwidth]{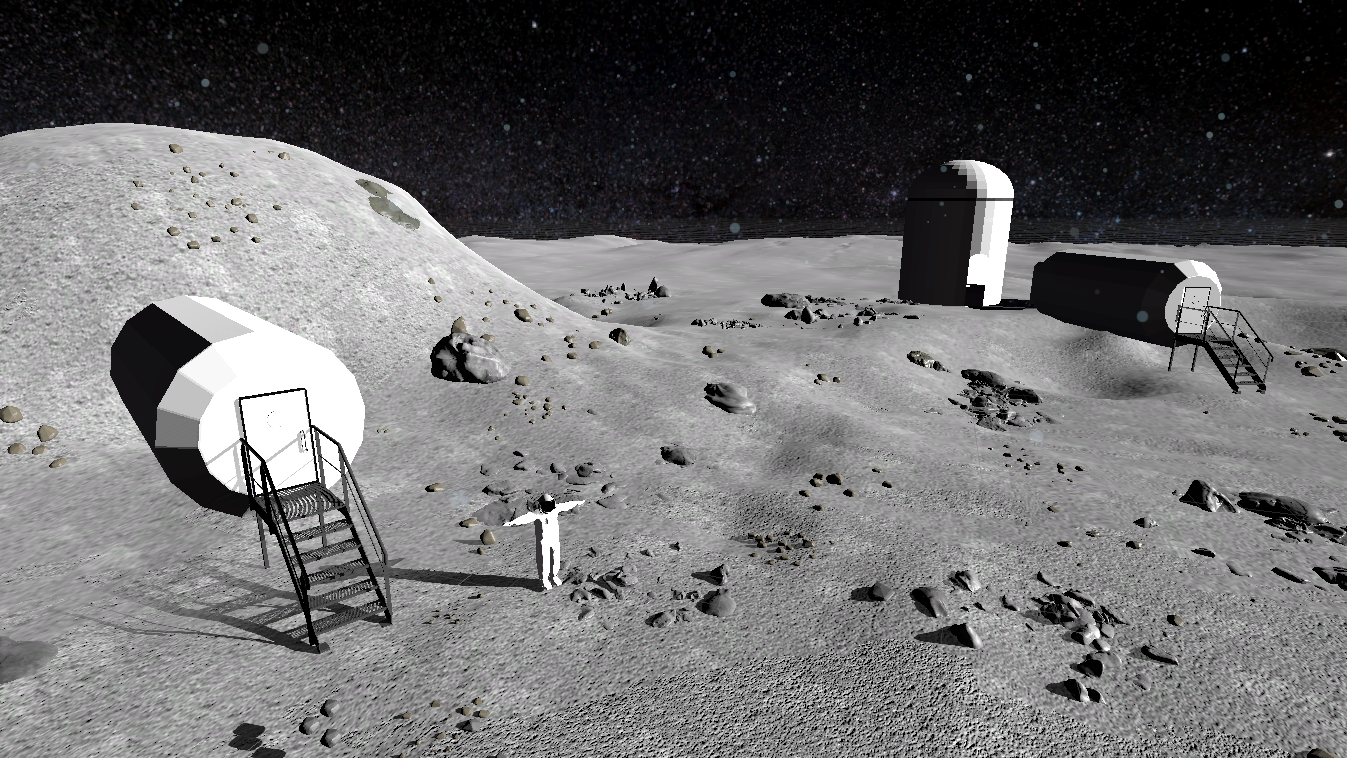}
\caption{South pole lunar base simulated in Unity.}
\label{fig:project1-overview}
\end{figure}

In this scene, a lunar base concept has been modeled. Terrain, rocks, and craters on the site have been reproduced based on available data and pictures mainly from Apollo missions, while two habitation modules have been designed based on Artemis base camp concepts.   
Special attention has been given to designing material textures to ensure they look realistic under the scene lighting conditions. To faithfully reproduce the latter, we made use of two components: global scene lighting and directional light. As background global illumination, we employed a Unity skybox: this asset consists of an HDRi, a 360-degree image with high dynamic range, which allows us to replicate the photographic lighting of an environment, in this case coming from the stars. We then modeled the Sun as a singular directional light, which is easy to configure and allows adjustment of intensity, angle, and elevation relative to the ground, enabling us to reproduce conditions of any desired site on the Moon. An overview of the scene is shown in Figure \ref{fig:project1-overview}.

This simulation has been developed to familiarize the  user with the environment in which, during a future long-term mission on the Moon, everyday activities will take place. Other than having realistic visual feedback, the astronaut can understand how to safely interact with objects when operating with a different gravity level compared to the one he is used to. In fact, it is easy to simulate objects' behavior under the influence of Moon gravity, thanks to Unity built-in physics engine.

\begin{figure}[b!]
    \centering
    \setlength{\fboxsep}{0pt}
    \centering
    \resizebox{\columnwidth}{!}{%
    \begin{tabular}{c c}
    \includegraphics[width=4.8cm, height=3.95cm]{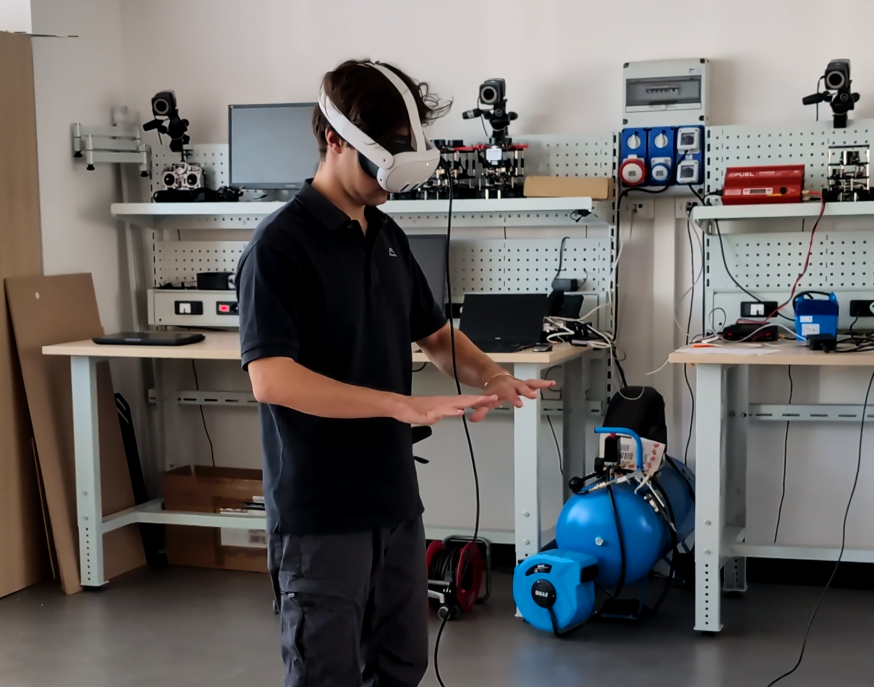} & \includegraphics[width=4.8cm, height=3.95cm]{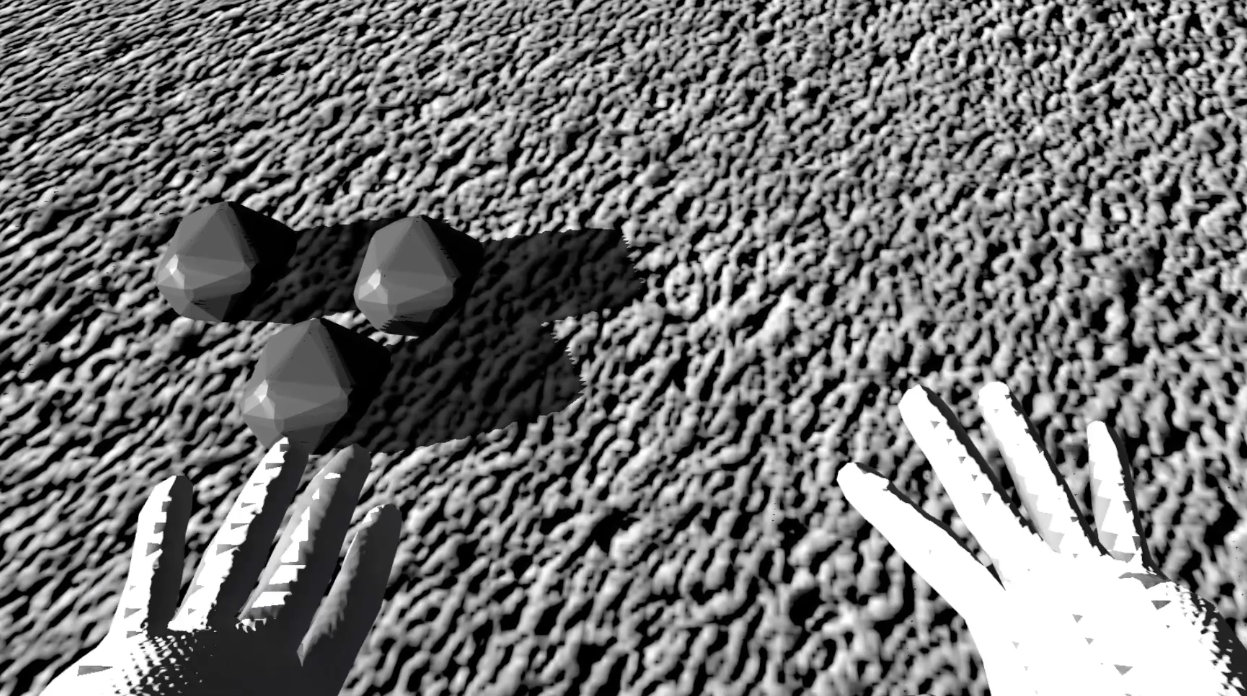} \\
    (a) & (b)
    \end{tabular}
    }
    \caption{Mapping of real world movements into simulation: user (a) and rigged avatar (b).}
    \label{fig:rigging}
\end{figure}

Figure \ref{fig:rigging} illustrates an example of real world user movements being mapped into the simulation avatar. To enable this feature, the avatar has been properly rigged: in the context of 3D modeling, the rigging process consists of the creation of a skeleton for the character, so that it can be animated according to input movements. An animation will in fact be associated with the bones components, which once activated will also move the associated mesh of the character accordingly. In our cases, rigging is necessary to perform proper inverse kinematics from headset measures to avatar movements. We leverage Meta SDKs to take advantage of headset external cameras to carry out precise hand and leg tracking, allowing to reproduce their movements in real-time inside the VR environment.

\subsection{VR driven mission design}
A key feature of our simulation framework is the capability to repeatedly test different mission tasks, while at the same time collecting useful insights from them. This is fully exploited thanks to the immersive experience provided by the VR headset. Specifically, while an astronaut is training, it is possible to understand how different designs of procedures and tools influence the operations, relying on the user's ability to correctly conclude the task, and iteratively update their properties in order to proceed with a new test campaign. 

\begin{figure}[!b]
\centering
\includegraphics[width=\columnwidth, height=7cm]{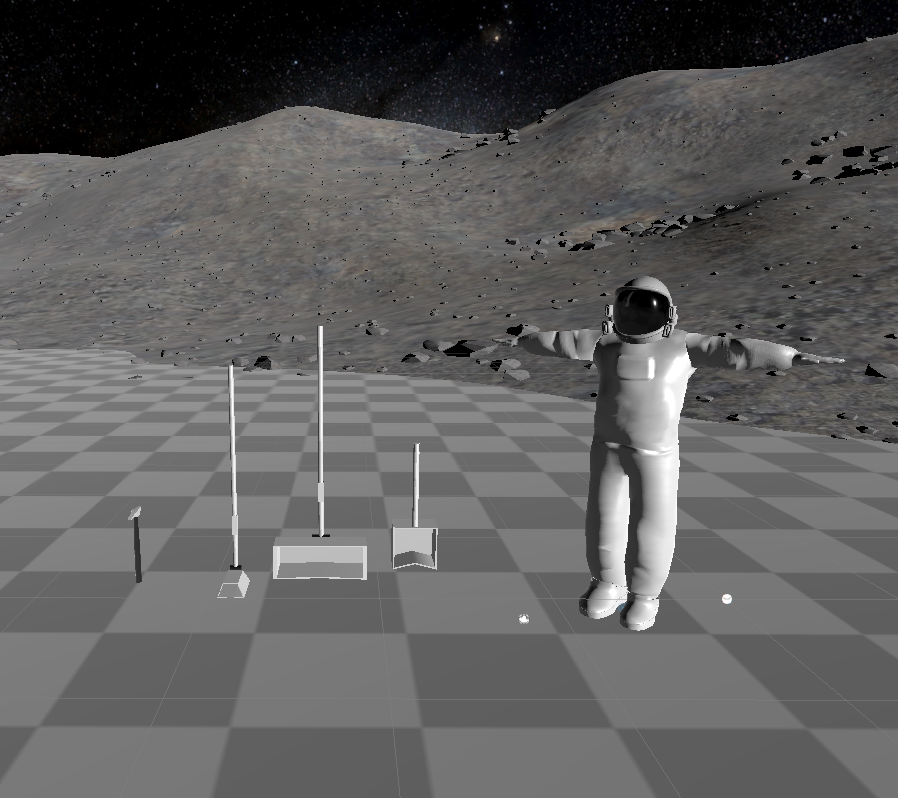}
\caption{Sample collection scene. Tools and avatar are placed on dynamic terrain texture rendered at runtime.}
\label{fig:vanghe}
\end{figure}

To illustrate this idea, we set up a scene to perform rock sample collection as shown in Figure \ref{fig:vanghe}. We modeled a series of tools for digging and shoveling with different characteristics, such as handle length and blade shape, and we assigned a specific rock sample to each one of them. The user can grab a tool, use it to interact with the terrain and learn how to collect and store samples. To visually give the idea of digging, a selected terrain area is provided with a CustomRenderTexture. This is a Unity object that allows to directly update a texture using a Unity Shader, for later use in a Material. Specifically, we perform ray casting from any game object which is active on the selected area (in our case, the astronaut avatar and the excavation tools) towards the terrain, that is voluntarily placed some centimeters under the texture. Results from ray casting are used to update the Shader that will generate a new material and render it on the CustomRenderTexture at runtime. 

\begin{figure*}[]
    \centering
    \begin{subfigure}[]{0.45\textwidth}
        \centering
        \includegraphics[height=5cm, width=8cm]{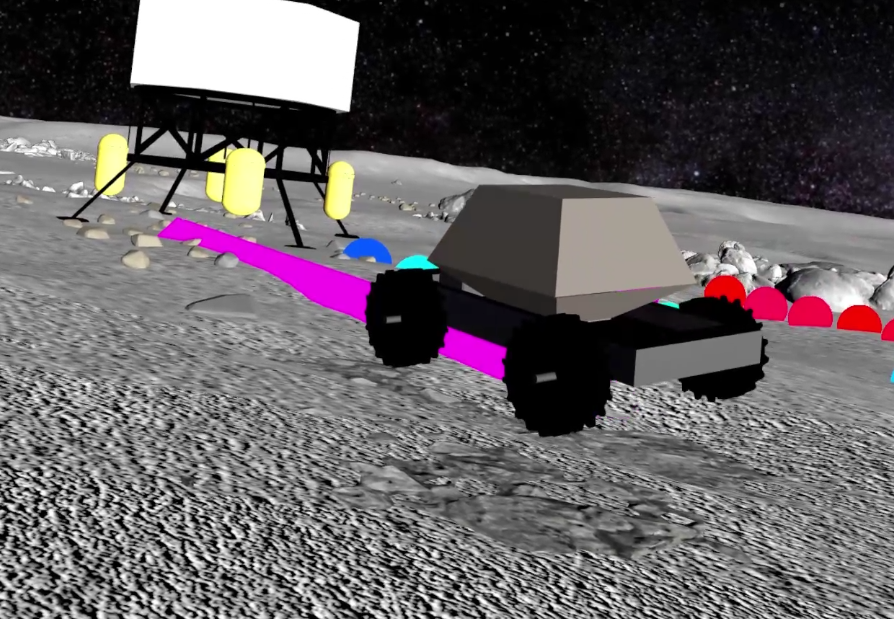}
        \caption{Unity simulation}
        \label{fig:sub1}
    \end{subfigure}
    \hfill
    \begin{subfigure}[]{0.45\textwidth}
        \centering
        \includegraphics[height=5cm]{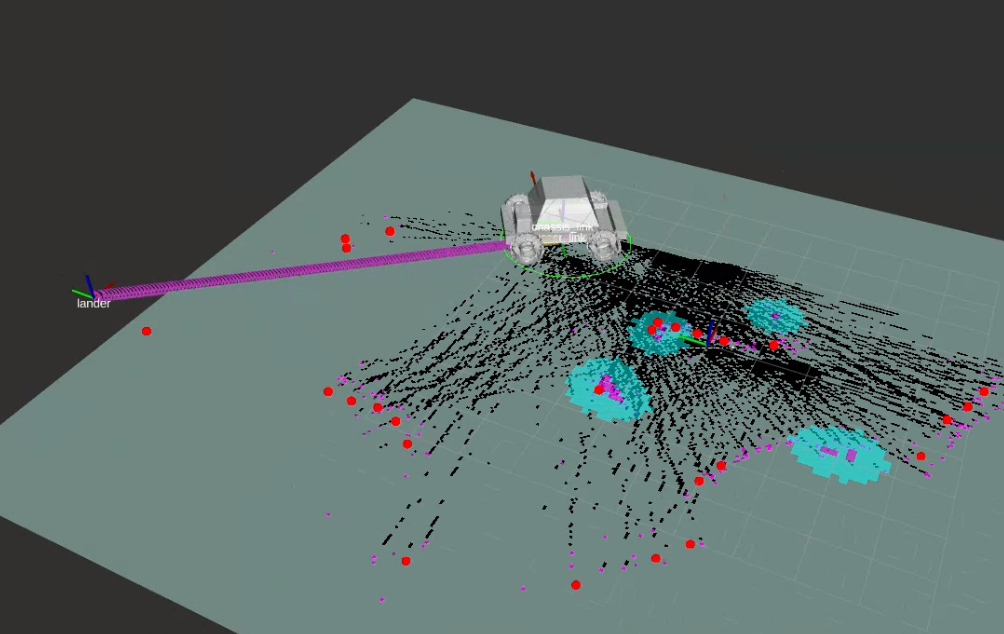}
        \caption{Rviz2 visualization}
        \label{fig:sub2}
    \end{subfigure}
    \caption{An example of Human-robot collaboration inside our framework. (a) robot autonomous behaviour is simulated in Unity, and (b) robot model and sensed environment are visualized in Rviz2. In both, markers for planned path and obstacles are shown.}
    \label{fig:ros-integration}
\end{figure*}

In this context, we expect that simulations will help in collecting valuable feedback from users' perspectives, which will be fundamental to understanding how to improve tools' properties, for example adjusting the blade shape to better hold the rock or the handle size in order to decrease required movements amplitude to accomplish the task.

\subsection{Robotic exploration}

In future long-term missions on the Moon, a tight integration between human operators and robotic agents is expected. As it is on Earth, robots can be applied to the execution of laborious, repetitive, or hazardous tasks, leaving astronauts with higher-level mission planning. Therefore, we believe that is important for upcoming astronauts to learn how to safely interact and collaborate with robots in a mission scenario.

In this context, we configured in our simulation framework a scene to replicate a typical scenario found during semi-autonomous robotic exploration of an unknown lunar region. To this end, we modeled an example of a Moon landing site near the South Pole: it contains the astronaut avatar and concepts of a descend lander and a semi-autonomous exploration rover. All the assets have been modeled in Blender, and the rover URDF is generated using the add-on Phobos \cite{phobos}. Once imported in Unity simulations, interfaces with the rover are exposed as ROS2 topics thanks to the plugins described in Section \ref{sub:Human-robot collaboration}. On the ROS2 side, we set up an environment for rover navigation based on the popular packages Navigation2 \cite{nav2} and Slam Toolbox \cite{slam_toolbox}. Here, the robot is configured to be able to autonomously navigate in a previously unknown environment, generating a 2D occupancy grid map from laser scan sensor readings, and planning a path over it to reach the desired goal set by the user. If needed, the rover can also be teleoperated with a joystick.
During navigation, the operator can always query the rover cameras to monitor the surrounding area from its point of view and, in any case, command to back up to a configurable base station. We also inserted to possibility for the user to be followed by the rover. An example of collaboration is illustrated in Figure \ref{fig:ros-integration}: here the user commanded the rover to move towards the lander.

To facilitate human understanding and interaction with the autonomous agent, the simulation includes a suite of tools designed to visualize real-time data processed by the rover. These are included in the Unity Robotics Hub Visualization Suite and provide insights into the rover's perception of the environment, including its sensor readings, generated map, and path planning decisions. By displaying this information in an intuitive manner, users can monitor the rover’s progress, diagnose potential issues, and intervene when necessary. Moreover, the visualization markers help to increase the security level of the operating environment, letting the user be aware of autonomous robots expected behaviour. 

\section{Results and Discussion}
A test campaign with non-professional subjects has been carried out in our laboratory to gain insights about the developed simulation framework. Tests have been focused on understanding the proposed solution ease of use, its efficiency as a training setup, and its limitations. Here we summarize the obtained results:
\begin{itemize}
  \item Simulation developed in Unity proved to be sufficiently realistic and able to replicate the lunar environment, on both visual and physical sides. Moreover, when inside a VR simulation, the user experience is fluent and commands are responsive. No appreciable lags with real movements are present.
  \item The framework proved to be generic and modular: expansion of a project and creation of new scenarios or use cases is simple, once the new assets have been modeled as desired. Their inclusion and setup in the simulations are straightforward. 
  \item Experiments also bring up limitations of the project, mostly related to collision managing performed by Unity. This is more appreciable when dealing with complex mesh colliders such as the one present on the astronauts' spacesuit, when it is possible to have a situation where two bodies collide with each other.
\end{itemize}

Building on the insights gained from our initial test campaign, future work will focus on two primary objectives. 
First, finding a trade off between realistic collision detection and computational complexity will be a key factor in improving our framework, giving more realistic behaviour of interactions between the user and other assets inside the simulation.
Secondly, we aim to extend the framework to additional use cases, allowing for broader applicability in various space mission simulations and training setups.

\section{Conclusions}
In this paper, a comprehensive framework for astronauts' training and mission design based on realistic VR simulations is presented. Key features of the framework have been illustrated and described with three typical use cases found in a future crewed mission to the Moon. Tests have been carried out, proving the ability of the user to exploit the simulations to perform mission operations and collaborate with robotic agents.    

\section*{Acknowledgements}
This publication is part of a PNRR PhD scholarship founded by MUR – DM 352/2022.

\bibliography{my_biblio}

\begin{thebibliography}{10}

\bibitem{10503238}
Vanshika Garg, Vaishnavi Singh, and Lav Soni.
\newblock Preparing for space: How virtual reality is revolutionizing astronaut training.
\newblock In {\em 2024 IEEE International Conference for Women in Innovation, Technology \& Entrepreneurship (ICWITE)}, pages 78--84, 2024.

\bibitem{Garcia}
Angelica~D. Garcia, Jonathan Schlueter, and Eddie Paddock.
\newblock {\em Training Astronauts using Hardware-in-the-Loop Simulations and Virtual Reality}.

\bibitem{Testi}
Corrado Testi, David~Z. Nagy, Joshua~E. Dow, and Olga Bannova.
\newblock {\em Use Virtual Reality as a Tool for Evaluating a Lunar Habitat}.

\bibitem{Nilsson}
Tommy Nilsson, Flavie Rometsch, Leonie Becker, Florian Dufresne, Paul Demedeiros, Enrico Guerra, Andrea Emanuele~Maria Casini, Anna Vock, Florian Gaeremynck, and Aidan Cowley.
\newblock Using virtual reality to shape humanity’s return to the moon: Key takeaways from a design study.
\newblock In {\em Proceedings of the 2023 CHI Conference on Human Factors in Computing Systems}, CHI '23, New York, NY, USA, 2023. Association for Computing Machinery.

\bibitem{McHenry}
Neil McHenry, Lauren Brady, Jaime Vives-Cortes, Erin Cana, Israel Gomez, Manuel Carrera, Kevin Mayorga, Javid Mustafa, Gregory Chamitoff, and Ana Diaz-Artiles.
\newblock Adaptive navigation for lunar surface operations using deep learning and holographic telepresence.
\newblock In {\em 2022 IEEE Aerospace Conference (AERO)}, pages 1--8, 2022.

\bibitem{curlin2022virtualrealitydigitaltwin}
Phaedra~S. Curlin, Madaline~A. Muniz, Mason~M. Bell, Alexis~A. Muniz, and Jack~O. Burns.
\newblock Virtual reality digital twin and environment for troubleshooting lunar-based infrastructure assembly failures, 2022.

\bibitem{unity}
{Unity Technologies}.
\newblock Unity, 2021.
\newblock Game development platform.

\bibitem{blender}
{Blender Development Team}.
\newblock Blender, 2023.
\newblock Computer software.

\bibitem{meta_xr_sdk}
Inc. Meta~Platforms.
\newblock Meta xr all-in-one sdk, 2024.
\newblock \url{https://developers.meta.com/horizon/downloads/package/meta-xr-sdk-all-in-one-upm}.

\bibitem{meta_movements_sdk}
Inc. Meta~Platforms.
\newblock Meta movement sdk, 2024.
\newblock \url{https://developers.meta.com/horizon/documentation/unity/move-overview/}.

\bibitem{ROS2}
Steven Macenski, Tully Foote, Brian Gerkey, Chris Lalancette, and William Woodall.
\newblock Robot operating system 2: Design, architecture, and uses in the wild.
\newblock {\em Science Robotics}, 7(66):eabm6074, 2022.

\bibitem{unity_ros2}
Unity Technologies.
\newblock Unity robotics hub, 2024.
\newblock Robotic simulation in Unity.

\bibitem{phobos}
Kai von Szadkowski and Simon Reichel.
\newblock Phobos: A tool for creating complex robot models.
\newblock {\em Journal of Open Source Software}, 5(45):1326, 2020.

\bibitem{nav2}
Steve Macenski, Francisco Martín, Ruffin White, and Jonatan Ginés~Clavero.
\newblock The marathon 2: A navigation system.
\newblock In {\em 2020 IEEE/RSJ International Conference on Intelligent Robots and Systems (IROS)}, 2020.

\bibitem{slam_toolbox}
Steve Macenski and Ivona Jambrecic.
\newblock Slam toolbox: Slam for the dynamic world.
\newblock {\em Journal of Open Source Software}, 6(61):2783, 2021.

\end{thebibliography}


\end{document}